\documentclass[%
 reprint,
 amsmath,amssymb,
 aps,
]{revtex4-1}

\usepackage{graphicx}
\usepackage{dcolumn}
\usepackage{bm}

\begin{document}

\title{Intruder penetration into granular matter studied by Lock-in Accelerometry}

\author{L. Alonso-Llanes} \author{G. S\'{a}nchez-Colina}\author{E. Mart\'{i}nez}
\affiliation{Group of Complex Systems and Statistical Physics, Physics Faculty, University of Havana, 10400 Havana, Cuba.}

\author{A. J. Batista-Leyva}
\affiliation{Instituto Superior de Tecnolog\'{i}as y Ciencias Aplicadas, 10400 La Habana, Cuba.}

\author{R. Toussaint}
\affiliation{Institut de Physique du Globe de Strasbourg (IPGS), Ecole et Observatoire des Sciences de la Terre (EOST), University of Strasbourg/CNRS, 67084 Strasbourg, France.}

\author{E. Altshuler}\thanks{corresponding author: ealtshuler@fisica.uh.cu}%
\affiliation{Group of Complex Systems and Statistical Physics, Physics Faculty, University of Havana, 10400 Havana, Cuba.}

\begin{abstract}
Understanding the penetration dynamics of intruders in granular beds is relevant not only for fundamental physics, but also for geophysical processes and construction on sediments or granular soils in areas potentially affected by earthquakes. In this work, we use Lock-in accelerometry to study the penetration of intruders into  quasi-2D granular matter fluidized by lateral shaking. We observed that there are two well-defined stages in the penetration dynamics as the intruder sinks into the granular material.

\begin{description}
\item[PACS numbers]
45.70.-n, 45.70.Mg, 07.07.Df, 07.50.Qx. 
\end{description}
\end{abstract}

\maketitle

By applying an external oscilatory force it possible to find a transition from a solid phase to a liquid phase in a granular media \cite{Duran}. Such effect produces the loss of solidity of the material and causes that an object laying on its surface sinks, tilts, or shifts laterally \cite{Duran}. This fluidization has a destructive effect during earthquakes because the buildings on its surface loss stability and eventually collapse \cite{ambraseys, liquefaction, manga}. Due to that earthquakes are one of the most destructive natural hazards to the social and economic structures of man. Although they can not be forecasted, their magnitude and after-effects can be minimized \cite{ambraseys}. In this paper we propose an experimental techinque aimed at understanding the effects of fluidization in order to reach those objetives.

While in quasi-2D systems the penetration of an intruder can be followed by means of a video camera \cite{dmaza, renaud, renaud2}, in the case of 3D systems of non-transparents grains this is not possible. Wireless accelerometry has been used in a few ocasions to quantify, as far as we know, the sink dynamics of an intruder \cite{goldman, seguin2, umba, pacheco}. In this note we use a method called Lock-in Accelerometry (LIA), previously reported by our group in order to circumvent those problems \cite{rsi}. As a result, we are able to establish two well-defined stages in the dynamics of penetration of an intruder into shaken granular matter, as the penetration depth increases.

Figure 1 shows the experimental setup which consists in a Hele-Shaw cell with a gap of $21.4 \pm 0.2 mm$ filled up with polidisperse spherical particles with $0.7 \pm 0.1 mm$ of average size and effective density of $0.715 g/cm^3$. At the bottom of the cell there is a horizontal hose with 30 holes of $0.5 mm$ diameter each, through which is possible inject air into the granular system. The cell is able to oscillate laterally using an electromagnetic shaker with an amplitude of $1.5 cm$ and a maximum frequency of  $\nu=6 Hz$. Released on the granular system there is an intruder, a squared parallelepiped $40 \pm 0.3 mm$ side, $17 \pm 0.3 mm$ thickness, and a weight of $51 \pm 1 g$, whose sinking is followed by a digital camera \textit{Hero2} that moves synchronously with the Hele-Shaw cell. Finally two 3-axis accelerometers MMA7456L are fixed to the Hele-Shaw cell (labeled Ref) and inside of the intruder (labeled Probe) \cite{AC}.

\begin{figure}
\includegraphics[width=8.5cm,height=6.5cm]{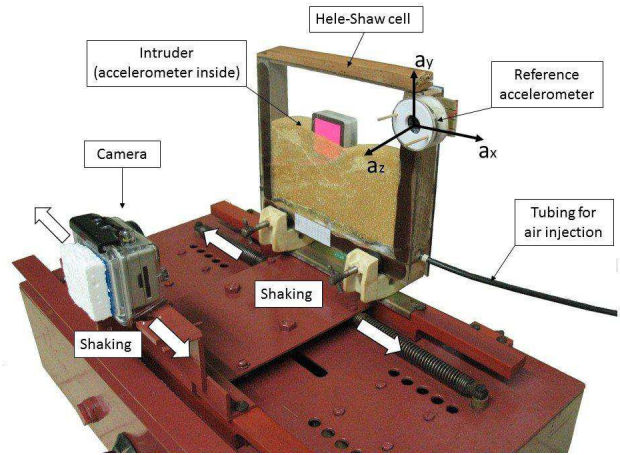}
\caption{\label{FIG} Experimental setup for quasi-2D measurements. Both the Hele-Shaw cell and the camera are synchronously shaken in the lateral direction. Accelerometers attached to the Hele-Shaw cell and the intruder bring the key information to quantify the sink dynamics}
\end{figure}

A typical experiment consists in injecting air into the granular system for 10 seconds ensuring the same initial conditions. Later is released the intruder (with the Probe accelerometer inside) on the free surface of the bed. Then is activated the data acquisition from the camera and accelerometers and lastly the electromagnetic shaker and the air injection systems are started at the same time until the sinking process ends.

The essence of the technique \textit{LIA} (LockIn Accelerometry) \cite{rsi} is the combined use of the information from the accelerometer fixed to the Hele-Shaw cell (Ref) and from the accelerometer inside of the intruder (Probe). Then, the experimental parameter used to study the sinking process is the correlation between the horizontal acceleretions from both accelerometers. For that it is used a modification of the Pearson's correlation coefficient aimed at decreasing the noise in the output that consist  in calculating the evolution of the Pearson's coefficient within time intervals of size $D$, each one starting at moment $k$, as follows:

\begin{equation}\label{pearson_mod1}
r\left(k\right)=\frac{\sum_{i=k}^{k+D}a_{x,R}\left(i\right)a_{x,P}\left(i\right)}{\left[\sum_{i=k}^{k+D}\left(a_{x,R}\left(i\right)\right)^{2}\sum_{i=k}^{k+D}\left(a_{x,P}\left(i\right)\right)^{2}\right]^{\frac{1}{2}}}
\end{equation}

where $a_{x,R}$ and $a_{x,P}$ represent the horizontal accelerations of the Reference and the Probe, respectively. $i$ represents the sampled time index and $N$ is the total number of experimental data points ($k$ runs from 1 to $N - D$).

The key idea behind this technique is that when the intruder is sinking, it will be a delay between $a_{x,R}$ and $a_{x,P}$ and the correlation coefficient will be smaller than one. As depth goes increasing the correlation increases because the probe is reaching a jammed phase into the granular system. Finally the value of the correlation must reach a plateau close to 1 indicating the end of the sinking process, where the intruder is moving synchronously with the reference.

\begin{figure}
\includegraphics[width=8cm,height=8cm]{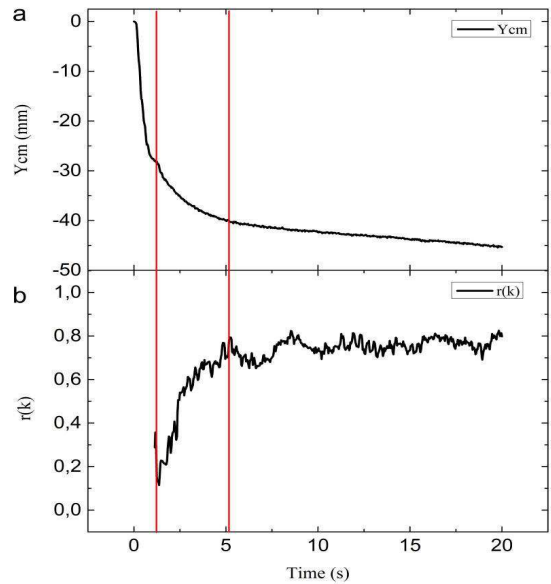}
\caption{\label{fig1} a) Penetration depth of the center of mass of the intruder versus time obtained from the video processing. b) Correlation coefficient versus time. With the help of the vertical guide lines it is easy to see that the changes in the correlation curve correspond to the changes in the sinking velocity of the intruder.}
\end{figure}

Figure \ref{fig1} shows the curves of a) the penetration depth of the center of mass and b) correlation coefficient versus time for an experiment with frecuency of $2.5Hz$ and air flux of $800 cm^3/h$. Figure \ref{fig1}a shows the penetration depth of the center of mass in time obtained by the image processing from the videos taken by the digital camera. The general behavior of the sinking may be characterized, fundamentally, by three stages: initially there is a fast sinking process, that takes around $1s$, then a decrease in the sinking velocity and after $5s$ a creep process takes place until the final stop. 

Figure \ref{fig1}b shows the time evolution of the correlation coefficient calculated using Eq. (\ref{pearson_mod1}) with $D = 30$. The first region in the correlation curve couldn't be measured due to technical limitations of the accelerometers. In $t=0$, as reported in \cite{rsi} for the case of the 3D experiment, the correlation should be 1 because both accelerometers are at rest. Then will have an initial fast-decrease in the correlation because in the initial moments, probe and reference has a delay in its horizontal accelerations as briefly discuss above. Later there is an increase of its value as a result of the arrival of the intruder to a region in the system in which it starts to move together with the granular mass. Finally the correlation remains at a saturation value until the end of the experiment.

\begin{figure}
\includegraphics[width=8cm,height=5.5cm]{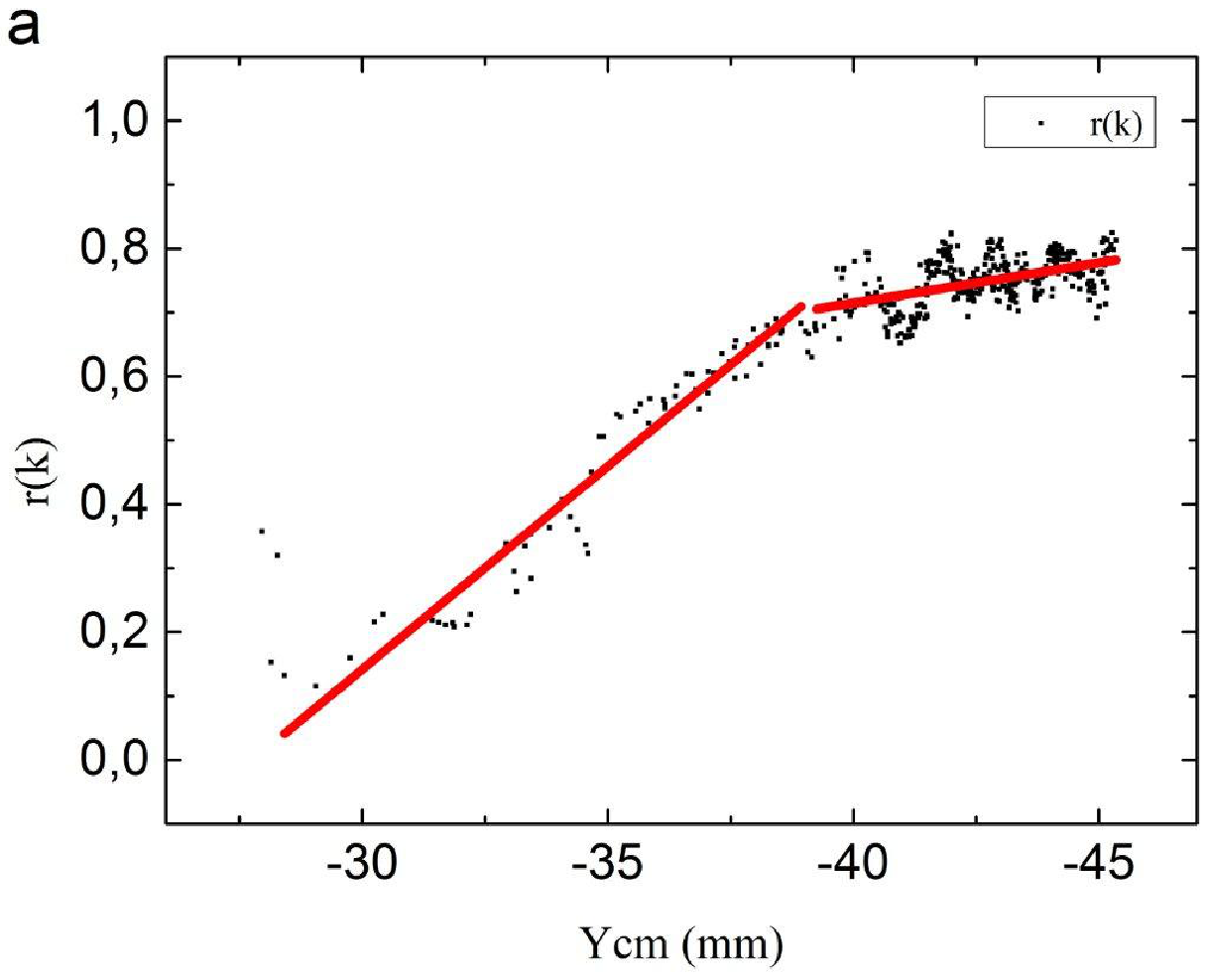}
\includegraphics[width=8cm,height=5.5cm]{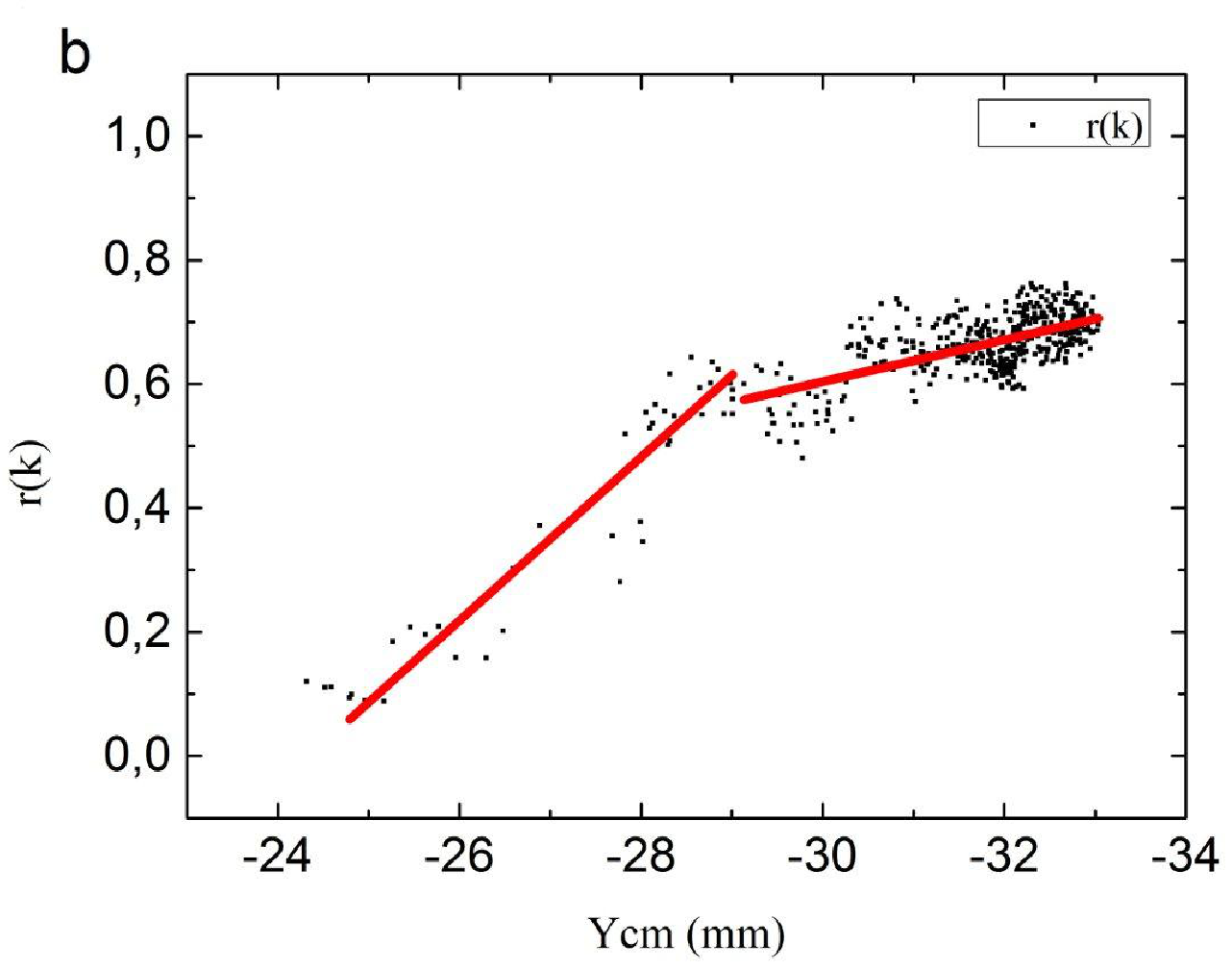}
\caption{\label{fig2} Correlation coefficient versus penetration depth of the center of mass a) for $2.5Hz$ and $800cm^3/h$ b) for $2.5Hz$ and $700cm^3/h$.}
\end{figure}

Figure \ref{fig2} shows the correlation coefficient as a function of the penetration depth of the center of mass for experiments with a) frecuency of $2.5Hz$ and air flux of $800cm^3/h$ and b) frecuency of $2.5Hz$ and air flux of $700cm^3/h$ where it is visible the influence of the air flux in the granular system. With the help of the guide lines it is possible to identify two main regions in each experiment, one between $29mm$ and $39mm$ of depth and the other from $39mm$ to $45mm$ of depth in Figure \ref{fig2}a and, in Figure \ref{fig2}b, one between $25mm$ and $29mm$ of depth and the other from $29mm$ to $33mm$ of depth. In the case of Figure \ref{fig2}a those regions match with the two regions after the first guide line in Figure \ref{fig1}. The previous curves represent the experimental verification of the technique's basis and may be useful as a calibration function to characterize the variation of the penetration depth of the center of mass of the intruder in the time from the temporal dependence of the correlation.

In the present contribution we have studied dry materials,
but the technique can also be used for wet granular matter. Finally, substituting the intruder by a solid rock and the granular bed by actual soil may expand the technique to measure,
in situ, the rheological response of a soil during an earthquake.

\end{document}